\documentclass[twocolumn,pre, superscriptaddress,floatfix,longbibliography]{revtex4-1}

\usepackage{graphicx,amsmath}
\usepackage{bm}
\usepackage{hyperref}

\usepackage{tabularx}
\usepackage{amssymb}
\usepackage{times}

\usepackage{color}
\usepackage{mathrsfs}

\hypersetup{
  colorlinks = True,
  urlcolor=blue,
  linkcolor = blue,
  citecolor= blue
}
\usepackage{tikz}
\usetikzlibrary{automata, positioning}
\usepackage{siunitx}
\usepackage{multirow}
\usepackage[capitalize]{cleveref}

\begin{document}

\title{Generation of  ice states through deep reinforcement learning}
\author{Kai-Wen Zhao}
\affiliation{Department of Physics and Center for Theoretical Physics, National Taiwan University, Taipei 10607, Taiwan}

\author{Wen-Han Kao}
\affiliation{Department of Physics and Center for Theoretical Physics, National Taiwan University, Taipei 10607, Taiwan}

\author{Kai-Hsin Wu}
\affiliation{Department of Physics and Center for Theoretical Physics, National Taiwan University, Taipei 10607, Taiwan}

\author{Ying-Jer Kao} 
\email{yjkao@phys.ntu.edu.tw}
\affiliation{Department of Physics and Center for Theoretical Physics, National Taiwan University, Taipei 10607, Taiwan}
\affiliation{National Center for Theoretical Sciences, National Tsing Hua University, Hsin-Chu 30013,  Taiwan}
\affiliation{Department of Physics, Boston University, 590 Commonwealth Avenue, Boston, Massachusetts 02215, USA}
\affiliation{Kavli Institute for Theoretical Physics, University of California, Santa Barbara, CA 93106-4030, USA}

\date{\today}

\begin{abstract}
We present a deep reinforcement learning  framework where a machine agent   is trained to search for  a policy to generate a ground state for the square ice model  by exploring the physical environment.
After training, the  agent is capable of proposing a sequence of local moves to achieve the goal.
Analysis of the trained policy and the state value function indicates that the ice rule  and loop-closing condition are learned without prior knowledge. 
We test the trained policy as a sampler in the Markov chain Monte Carlo and benchmark against  the baseline loop algorithm.
This framework can be  generalized to other models with topological constraints where generation of constraint-preserving states is difficult.
\end{abstract}

\maketitle

\section{Introduction}

Ice models~\cite{Lieb:1967jw} are the simplest models that can describe the statistical properties of  proton arrangement in water ice~\cite{Pauling:1935ay}, and spin configurations of spin ices~\cite{bramwell2001spin}.
The ground states in the ice model follow a local constraint, called the \textit{ice rule}.
Due to this constraint, performing local changes in a given ice configuration does not create a new ice state. 
The loop algorithm, widely used in the Monte Carlo sampling of ice models, generates ice states  by exploiting the fact  that the difference between two ice configurations are in the form of  closed loops~\cite{Rahman1972Stillinger,Yanagawa:1979jk,Barkema:1998eo}.
This is an example where  an efficient algorithm generating new constraint-satisfying states  emergences from clever inspection of sample states.
We want to explore how to automate this type of discovery with  machine learning techniques and provide a proof-of-principle implementation. 

Efficient state generation is important, in particular, in Markov chain Monte Carlo (MCMC)~\cite{landau2014guide, newman1999monte}. 
To generate statistically sound samples in the simulation, we need to reduce the autocorrelation between samples in the Markov chain.
This  can be achieved by clever design of update proposals; for example, in the case of two-dimensional ferromagnetic Ising model,  cluster updates such as Swendsen-Wang and Wolf algorithms~\cite{swendsen1987nonuniversal, wolff1989collective} are powerful methods to reduce the effects of critical slowing down near the critical point and allow for precise extraction of physics with larger systems.
However, these cluster algorithms are normally designed for specific models  and it is hard to generalize.

Recently, a new class of proposals termed self-learning Monte Carlo, which uses configurations generated by a small size simulation to train an effective model, and draws samples from the effective model to generate new configurations,  have  demonstrated success in improving of  the efficiency of both classical and quantum Monte Carlo simulations~\cite{liu2017self, nagai2017self, liu2017self2, xu2017xy}. 
However, the amount of training data required for a successful training normally scales with the complexity of the effective model.
Another class of proposals uses the generative models based on deep neural networks in order to generate new samples.
The restrict Boltzmann machine, a  neural network architecture connected to the the real-space renormalization group~\cite{Mehta:2014ua},
has been used to generate Monte Carlo samples to study thermodynamics~\cite{Torlai:2016kq} and to accelerate Monte Carlo simulations~\cite{Huang:2017sf}.
The generative adversarial network~\cite{Goodfellow:2014fv}, where two neural networks contesting with each other, is used to generate configurations for  two-dimensional Ising model~\cite{Liu:2017th} and for a complex scalar field in two dimensions with a finite chemical potential~\cite{Zhou:2018rw}.
The success behind these methods relies on the fact that the underlying distribution of the configurations is continuous in the thermodynamic limit; that is, all configurations generated by the neural networks are allowed. 
On the other hand, direct application of these methods to generate configurations for the topologically constrained model can be difficult.
Here not all  configurations generated by the neural network can satisfy the constraint. 
Therefore, a new scheme is highly coveted. 

In this paper, we explore how to apply a reinforcement learning framework~\cite{sutton2018reinforcement} to generate new configurations that satisfy the ice rule constraint.
Reinforcement learning has demonstrated remarkable abilities in achieving better than human performances on video games~\cite{mnih2015human}, and  the game of Go~\cite{silver2016mastering,silver2017mastering}.
 Recently it  has also been applied to  quantum physics research such as quantum control~\cite{Fosel:2018wc}, quantum error correction~\cite{Sweke:2018hs,Liu:2018xi,Poulsen-Nautrup:2018jy}, quantum experiment design~\cite{Melnikov:2018vq}.
In a reinforcement learning setup, a machine agent interacts with its environment. 
At each time step, the agent takes an action on the current state of the environment, and then receives a calculated reward from the observation of the consequence of this action.
 A feasible reinforcement learning algorithm seeks to maximize the total reward  through a trial-and-error learning procedure. 
We will utilize this feature to design an algorithm to adaptively search for a move from an ice state to another.%
Similar ideas have been explored recently by the policy-guided Monte Carlo (PGMC) proposal~\cite{pgmc2018} to  improve MCMC sampling.
The policies in Ref.~\cite{pgmc2018} are modeled with simple functions and the best policy is obtained by  optimizing parameters in these models. 
Here we instead model the policy as a deep neural network, and  build the physical constraints into the design of the reward function. 
Therefore, we can take advantage of the recent progress in the reinforcement learning research to train an agent with a simple set of actions to explore the physical environment. 
The  best policy naturally  emerges out of the agent's interaction with the environment, subjected to the reward design.  
This scheme should serve as a general framework to search for methods to generate topologically constrained states in other physical models.

The rest of the paper is organized as follows: In Sec.~\ref{squareicemodel}, we introduce the square ice model and the conventional loop algorithm.
 In Secs.~\ref{RL} and \ref{sec:agent}, we introduce the deep reinforcement learning framework we use.
 In Sec.~\ref{training}, the training process and agent behavior will be discussed.
 We use the trained policies as  samplers for Markov chain Monte Carlo in Sec.~\ref{sampler}. 
 Finally, we conclude in Sec.~\ref{conclusion}.

\begin{figure}[t]
  \centering
  \includegraphics[width=\columnwidth]{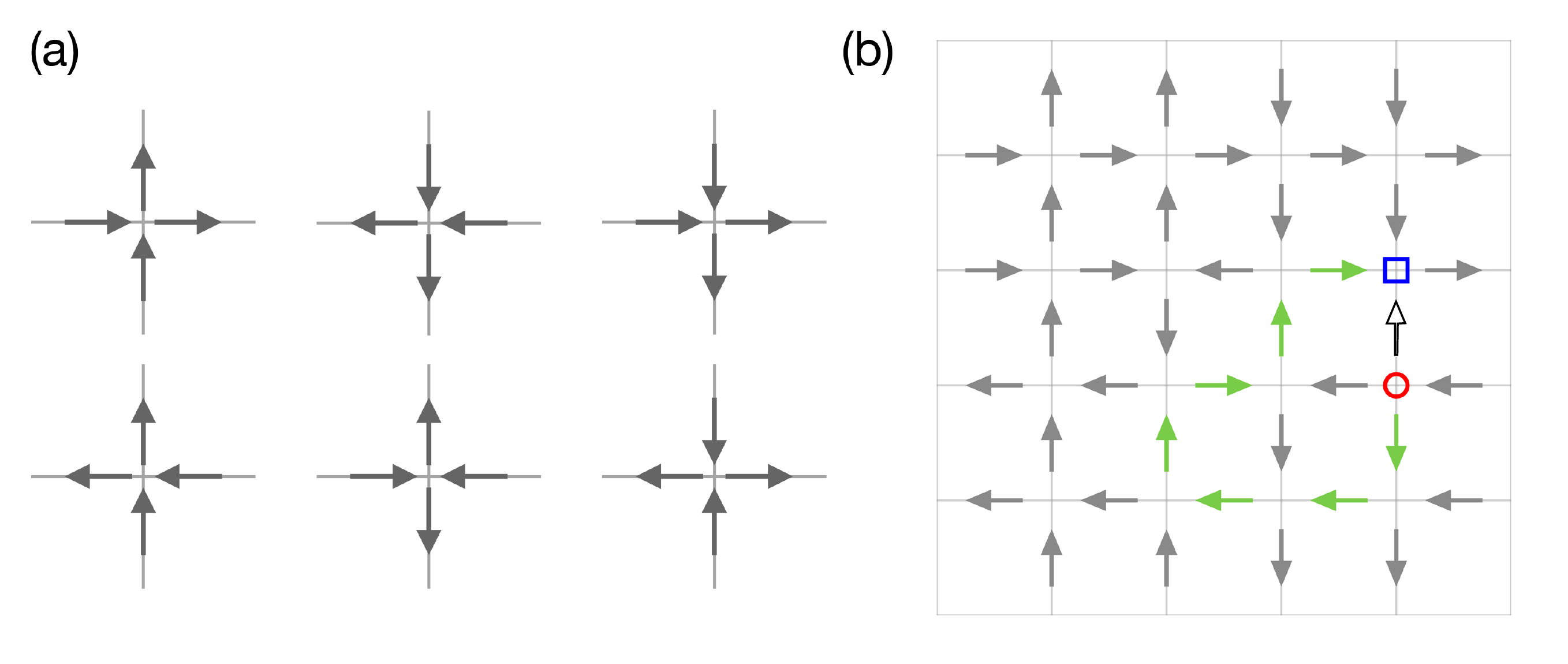}
  \caption{
    \label{fig:square_ice}
     (Color online) (a) Local spin configurations that satisfy the 2-in-2-out ice rule at each vertex. 
      (b) Flipping a spin (unfilled arrow) in an ice state creates a pair of defects, resulting in a high-energy state. 
      Propagation of a single defect (red circle) by flipping green (bright) spins creates no additional defects.
      When the loop is closed, the pair of defects (red circle and blue square) annihilates and  a new ice configuration is generated.
  }
\end{figure}

\section{Square Ice Model\label{squareicemodel}}

\begin{figure}
  \centering
  \includegraphics[width=\columnwidth]{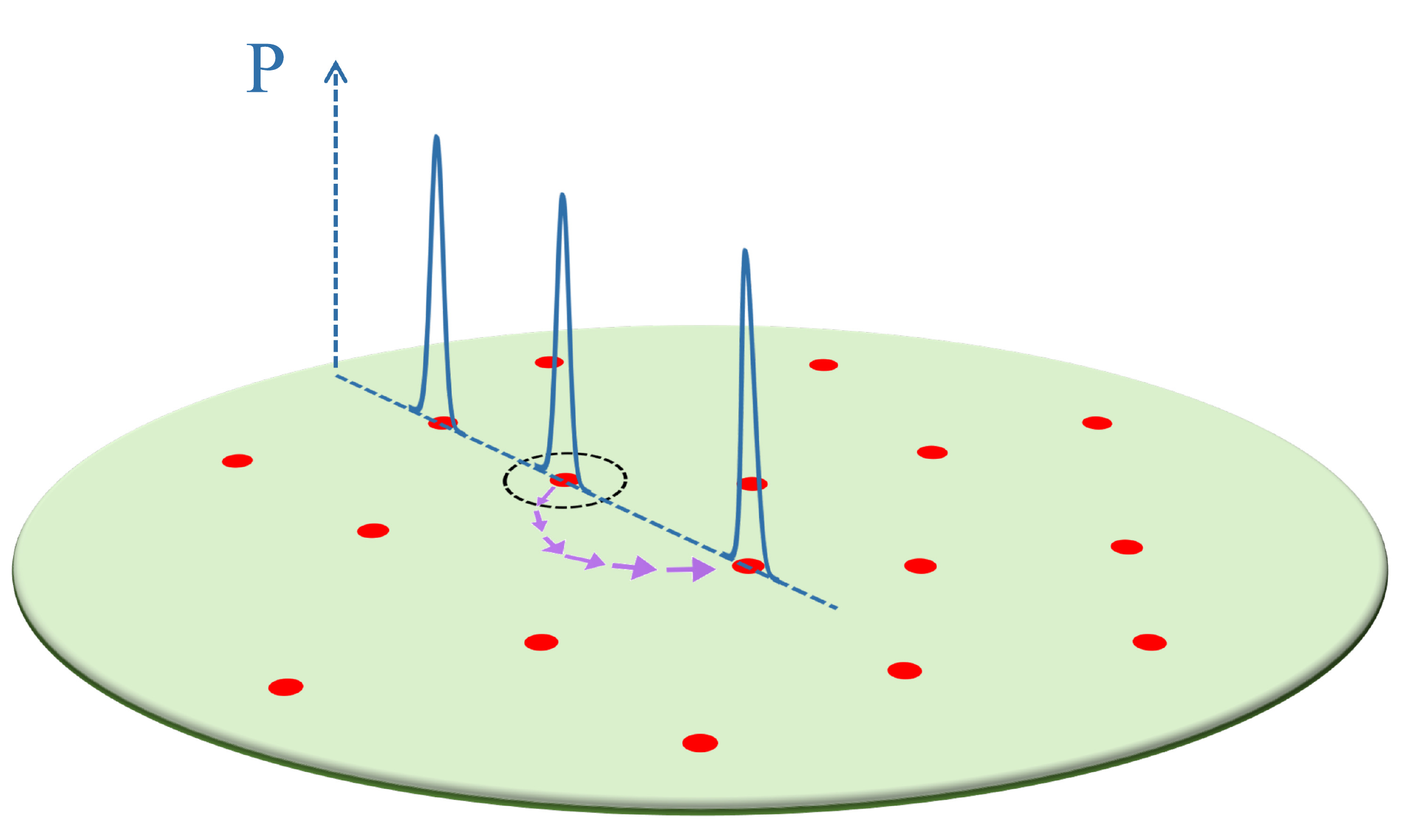}
  \caption{
    \label{fig:phasespace}
    (Color online) Ice configurations are sparsely populated in the full configuration space. 
    Each red dot represents an ice configuration. 
    The distance between the red dots measures the difference between  ice configurations.
    The black (dashed) circle indicates the configurations reachable by a single-spin flip.
The probability $P$ for each ice configuration is a $\delta$-function; therefore a single-spin flip alone can not reach another ice configuration.
By constructing a path (curved arrows) of single-spin flip moves, it is possible to reach  a new ice configuration. 
  }
\end{figure}

Here we consider a two-dimensional  ice model on a square lattice~\cite{Lieb:1967jw}. 
At each vertex, the arrangement of spins satisfies the ice rule: two spins pointing into and two pointing out of  the vertex~(Fig.~\ref{fig:square_ice}(a)).
There exist  six types of vertices that satisfy this constraint and all other types of vertices are considered defects with higher energy. 
This is the reason why ice models are sometimes referred as six-vertex models~\cite{Lieb:1967jw,Baxter:2016df}.
Ice states, therefore, correspond to spin configurations with zero defects. 
All ice states have the same energy, and  they   form an extensively degenerate ground state manifold with a finite residual entropy at zero temperature of  $S/N=k_B\ln W$ where $W=\left(4/3\right)^{3/2}$~\cite{Lieb:1967jw}.
They are, however, sparsely populated in the configuration space of all possible spin configurations~(\cref{fig:phasespace}).
Consider, for example, flipping a random spin from a given ice configuration, and the two neighboring vertices connected by this spin violates the ice rule constraint, creating a pair of topological defects of three-in-one-out and one-in-three-out vertices~(Fig.~\ref{fig:square_ice}(b)) .
The configurations that satisfy the ice rules are thus separated by large energy barriers and the conventional Metropolis single-spin flip update scheme will fail. 
However, it is possible to flip a series of neighboring spins to bring one ice configuration to another. 
By examining the difference between two ice states, it is clear that they differ only by spins forming closed-loops. 
This indicates  loop-like updates can  bypass the energy barrier and reach a new  ice configuration. 

The simplest algorithm to generate random closed loops is the long loop algorithm~\cite{Rahman1972Stillinger,Yanagawa:1979jk}. 
In each round of the algorithm, one vertex is  chosen as a starting point. 
One of the two out-going spins from this vertex is chosen with equal probability~\cite{Barkema:1998eo}. 
Following this spin to the connecting vertex and choose one of the  two out-going spins with equal probability. 
Repeat the process until the starting vertex is reached and the loop closes.
All the spins on the resulting loop are reversed to update the configuration while preserving the ice rule. 
Since at each vertex out-going spins are chosen with equal probability, there are  $2^{m}$ possible paths for a given starting point, where $m$ is the size of the loop. 
This algorithm gives an example of how one can generate a new ice configuration by going through a series of highly unfavorable configurations in order to achieve the final goal. 
This philosophy is indeed very similar to that of the reinforcement learning that the  best long-term strategy may involve short-term sacrifices. 
It is natural to explore the reinforcement learning scheme to generate new ice configurations.

\section{Reinforcement Learning\label{RL}}

Reinforcement learning (RL) is the branch of machine learning concerned with making sequence of decisions for long-term profit ~\cite{sutton2018reinforcement}. 
The central idea of RL is to consider a machine agent situated in an environment. 
At each time step, the agent takes an action on the current state of the environment, and then receives a calculated reward from the observation to the outcome of this action.
 A feasible RL algorithm seeks to maximize the total reward for the agent from an unknown environment through a trial-and-error learning procedure. 
We model the agent policy  $\pi_\theta$ with parameters $\theta$ using deep and convolutional neural networks. 

\begin{figure}
  \centering
  \includegraphics[width=\columnwidth]{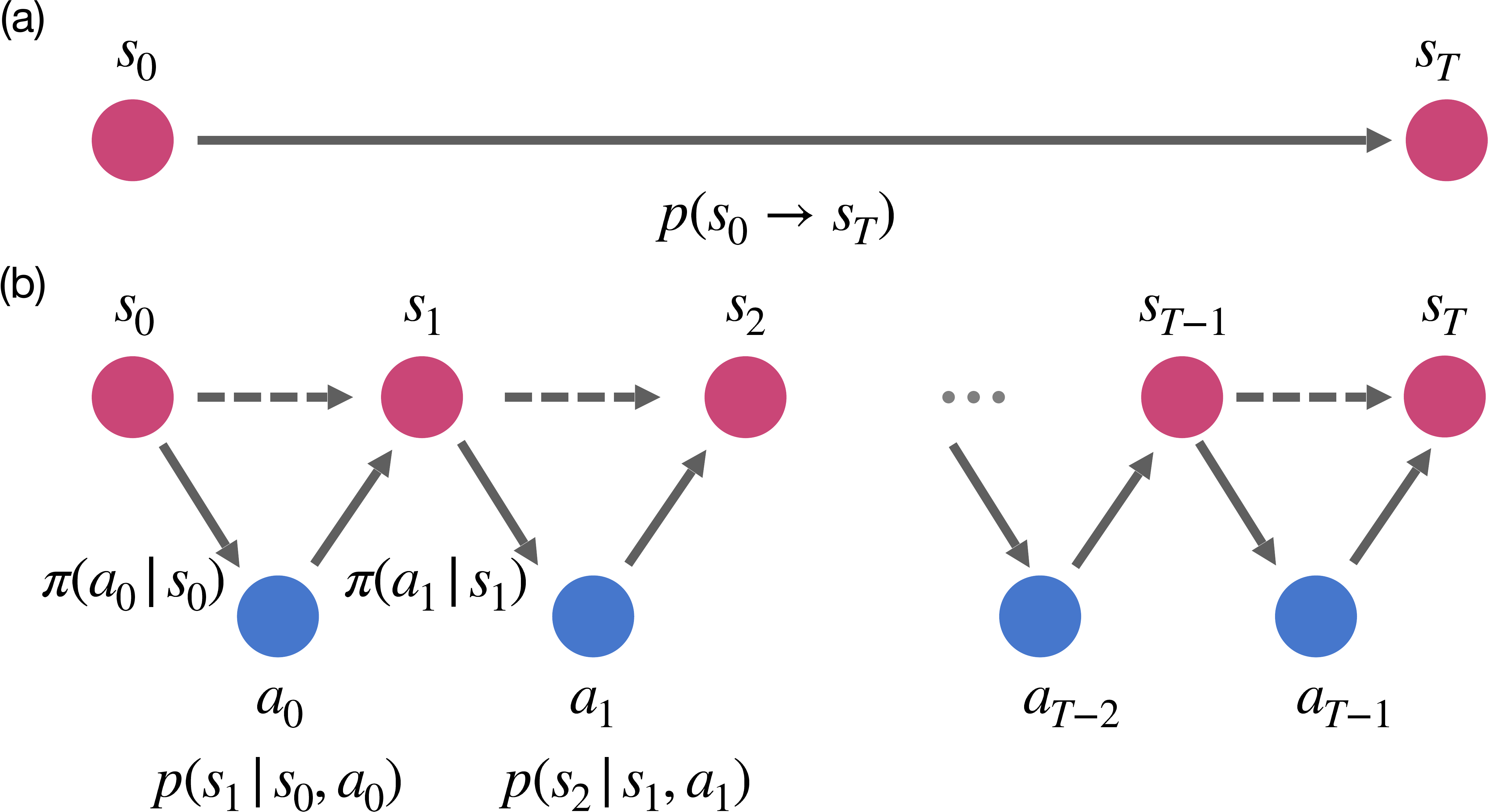}
  \caption{
    \label{fig:mdp}
    (Color online) A global move in the Markov chain (a) from state $s_0$ to $s_T$ can be decomposed into a sequence of local  moves.
    By extending the Markov chain to a Markov decision process, the policy $\pi(a|s)$ serves as a transition operator, making local decisions, to generate a global movement in state space. 
  }
\end{figure}

The core idea of our work is to parameterize the proposal operator as the agent policy and search for efficient transitions in the configuration space under given physical constraints.
In order to automate the search process, we extend the original Markov chain to a Markov decision process (MDP), by indicating how a state $s_t$ transitions into a new state $s_{t+1}$ using the action $a_t$  with the transition probability $p(s_{t+1} | s_t, a_t)$. 
The parametrized policy $\pi_\theta(a|s)$ specifies  the probability of the action $a$  the agent  will take given the state $s$.
A reward function $r_t\equiv r(s_t, a_t)$ is associated with each state-action pair $\{s_t, a_t\}$. 
A move in the Markov chain can, therefore, be decomposed into a series of decision-makings in the MDP (\cref{fig:mdp}),
\begin{equation} \label{eq:mcmdp}
  p(s_0 \to s_T) = p(s_0) \prod_{t=0}^{T-1} \sum_{a_t}\pi_\theta(a_t|s_t) p(s_{t+1} | s_t, a_t)
\end{equation}
The policy $\pi_\theta$ can be regarded as a proposal operator in the conventional MCMC. 

We briefly sketch the policy gradient scheme used in this work and refer the interested readers to Refs.~\cite{sutton2018reinforcement,Neftci:2019hc} for details. 

Given a trajectory of the state-action pairs $\tau=(\{s_0,a_0\},\{s_2, a_2\},\ldots)$  in the MDP, the goal of RL is to maximize the expected future return, \textit{i.e.}, the weighted sum of rewards. 
The objective function can be defined as a function of the policy, 
\begin{equation} \label{eq:obj_function}
  J(\pi_\theta) = \mathbb{E}_{\pi}[R(\tau)] =\sum_{\tau} \pi_\theta(\tau)R(\tau) ,
\end{equation}
where $\mathbb{E}_\pi$ denotes the expectation value is evaluated along a trajectory $\tau$ following the policy $\pi$. 
The total expected return along the trajectory $\tau$ is given as $R(\tau) = \sum_{t=0}^{T-1} \gamma^{t} r_{t}$, where $0\le\gamma<1$ denotes a discounting factor.
The general method for estimating the gradients of  an expectation function is using the score function estimator~\cite{Shapiro:2003ir}, and the model-free policy gradient can be obtained as
\begin{equation} \label{eq:policy_grad}
    \nabla_\theta J(\pi_\theta) = \mathbb{E}_{\pi}[ \nabla_\theta \log \pi_\theta (\tau)R(\tau)].
\end{equation}
The optimal parameter $\theta$ can be obtained by the gradient ascent $\theta \leftarrow \theta+\alpha\nabla J(\pi_\theta)$ with the learning rate $\alpha$. 

To estimate how good it is for the agent to be in state $s$,  following policy $\pi$, we  define a  state-value function, 
\begin{equation} \label{eq:value_fun}
    V^{\pi}(s) = \mathbb{E}_{\pi}\left [\left.\sum_{k=0}\gamma^{k}r_{t+k+1}  \right| s_t=s\right],
\end{equation}
which  estimates the expected future rewards starting from state $s$. 
%
%
%
%
To estimate $V^\pi$, one  starts from a random initial state and follow  $\pi(a|s)$ to generate a trajectory to 
accumulate average return achieved from each state. 
Given enough samples, the average should converge to the optimal $V_\pi$. 
The value function allows us to compare the policies: a policy $\pi$ is better than or equal to a policy $\pi'$ if only if $V_\pi(s)\ge V_{\pi'}(s)$ for all $s$~\cite{sutton2018reinforcement}. 

We use  the {Asynchronous Advantage Actor-Critic} (A3C) algorithm~\cite{mnih2016asynchronous} to perform the training.
The value function $V^\pi_\phi(s)$ is  also parametrized by a neural network  with parameter $\phi$, and the policy gradient is given by
\begin{equation}
\nabla_\theta J(\pi_\theta) = \mathbb{E}_\theta\left[ \sum_t \nabla_\theta \log \pi_\theta (a_t| s_t) A^\pi_{\phi}(s_t, a_t)\right],
\end{equation}
 where  $A^\pi_{\phi}(s_t,a_t) =r_t+\gamma V^\pi_\phi(s_{t+1})-V^\pi_\phi(s_t)$ is an estimate of the advantage function~\cite{mnih2016asynchronous}. 
 The parameterized value function $V^\pi_\phi (s)$ is trained to predict the expected future return that reduces the variance of policy gradient and stabilizes the training process.
In practice, the training is done using the  loss function,
\begin{widetext}
\begin{equation} \label{eq:acloss}
  L(\theta,\phi) = -\left[\sum_t \Big(\log\pi_{\theta}(a_t, s_t) A^\pi_{\phi}(t)
  - ||V^\pi_\phi (s_t)-R_t||^2 \Big)
  + \lambda\pi_{\theta}(a_t|s_t)\log\pi_{\theta}(a_t|s_t)\right].
\end{equation}
\end{widetext}
The first term describes the policy loss, the second term gives the estimation of the value function, and the last term provides the entropy regularization. 
In our experience that the policy entropy term not only promotes exploration but provides a sensible indicator in the training process.
The asynchronous algorithm launches several independent agents sampling
trajectories from distinct environments. 
A central network holds shared parameters and is updated asynchronously by many agents. 
In other words, each agent explores the state space simultaneously from different initial states. 
This approach prevents the machine from overfitting to some specific configuration and generalizes the policy search. 
Eight independent agents are used in our program to  update eight independent ice states simultaneously.

\section{Agent-Environment Interface\label{sec:agent}}

\begin{figure}[tb] 
  \centering
  \includegraphics[width=0.8\columnwidth]{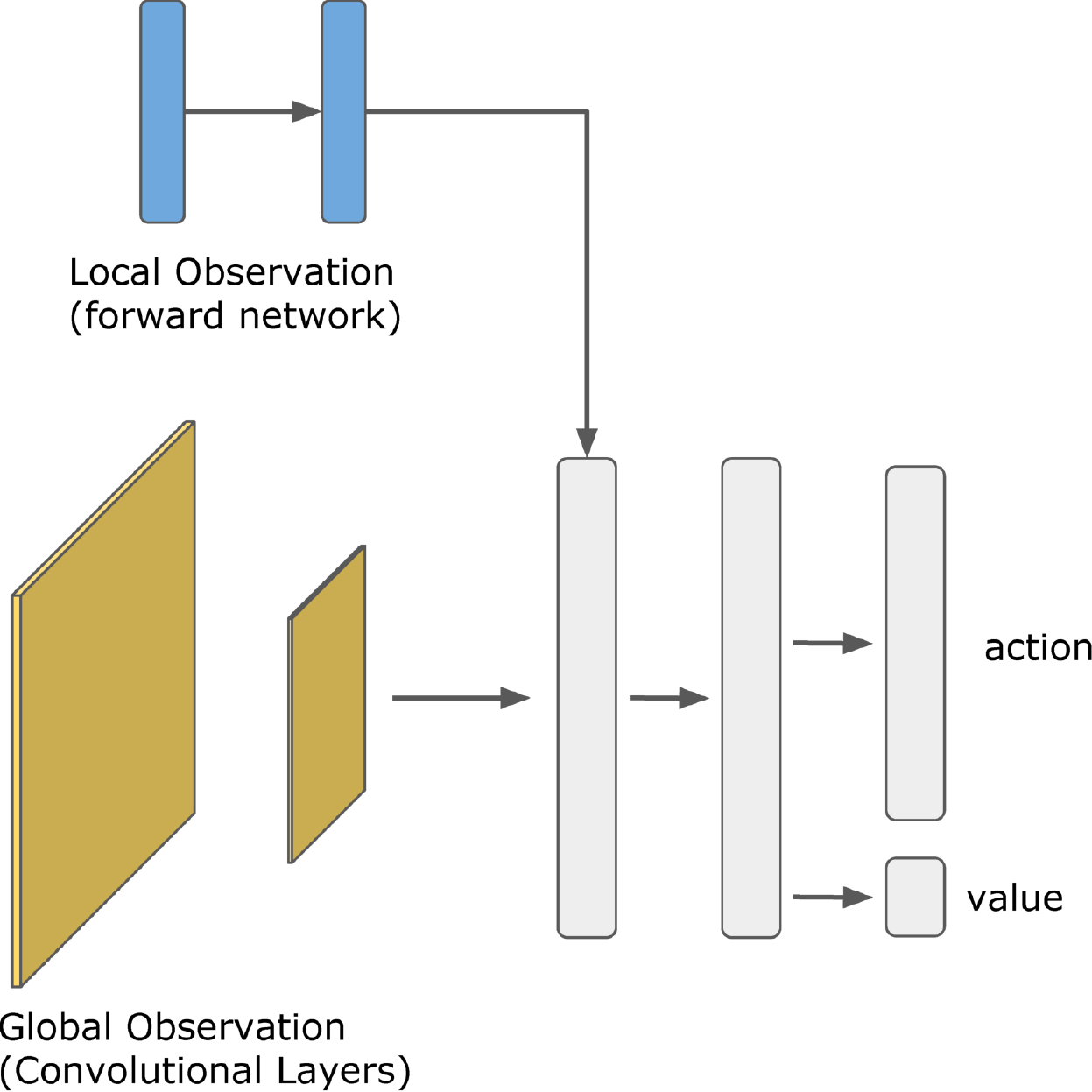}
  \caption{
    \label{fig:arch} (Color online)
    Architecture of the multi-channel neural network  for the agent.
    For the NN policy,  global observation is turned off.  }
\end{figure}

We implement an interactive interface for the interaction between the machine agent and the physical environment compatible with OpenAI  Gym \cite{brockman2016openai}. 
 Two kinds of observations of the environment are provided to the agent.
The global observation $O_{\rm global}$ monitors the configuration changes $C_t=s_t-s_0$ between the initial ice state $s_0$ and the current state $s_t$.
The local observation $O_{\rm local}$ monitors  local information such as the neighboring spins $\sigma_i$, the fraction of configuration change $\Delta C$ defined as the ratio of the number of flipped spins and the total number of spins. 
A state in the MDP is a composite of local and global observations, $s = [O_{\rm local}, O_{\rm global} ]$. 
In deep reinforcement learning, the policy and value functions are both approximated by neural networks.
 In practice, the value function usually shares weights with the policy approximator~\cite{vinyals2017starcraft}; therefore,  the same network is used for making action decision and predicting the future return.
 We model the agent with two types of networks: (a) Local network, where only  local observation is used as input information; (b) Multi-channel network, where local and global observations are combined. 
Figure~\ref{fig:arch} shows the architecture of the multi-channel neural networks for the agent~\cite{vinyals2017starcraft}.
The local information is fed into two layers of  feedforward  networks while the global observation is extracted through two layers of convolutional neural networks before they are concatenated.  
The specification of the neural network architecture and hyperparameters for training can be found in App.~\ref{app:network}.
In the following, we call the policy obtained from the local network as NN policy, and multi-channel network as CNN policy.

\begin{figure}[tb] 
  \centering
  \includegraphics[width=\columnwidth]{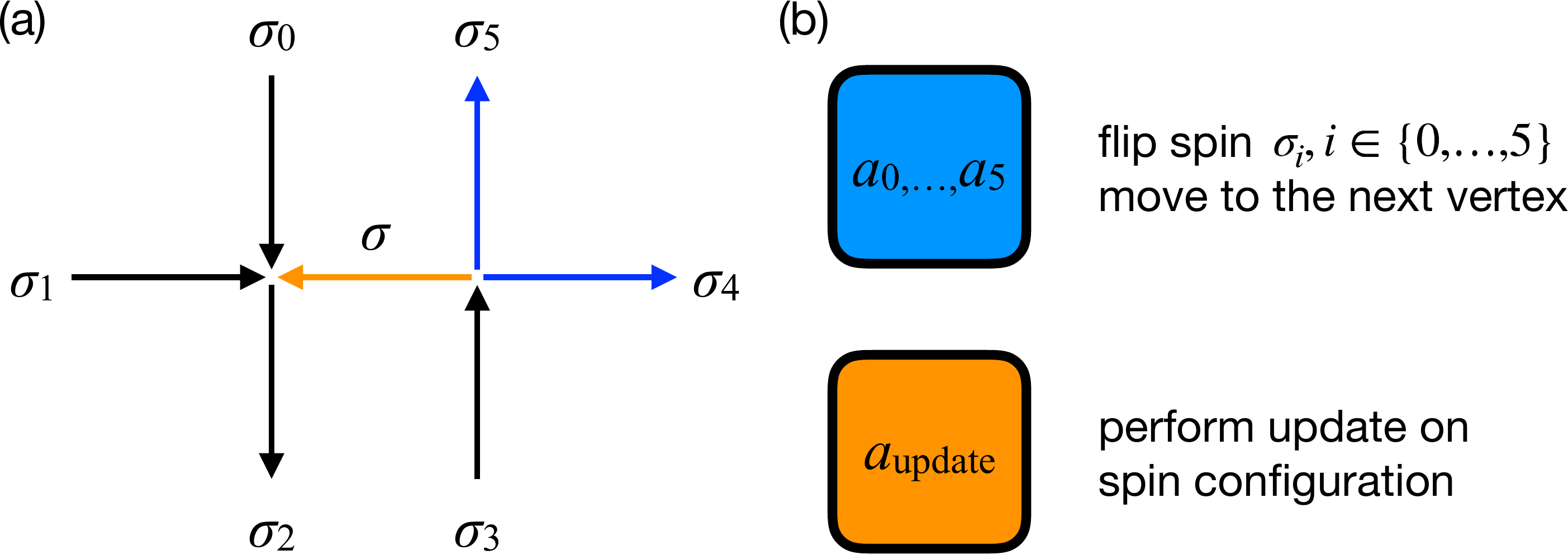}
  \caption{
    \label{fig:local} (Color online) (a) Local environment for an agent located at spin $\sigma$. 
    There are six neighboring spins that the agent can choose to flip next. 
    The blue spins $\sigma_4$ and $\sigma_5$ correspond to the possible choices in the loop algorithm.
    (b) Two types of actions that the agent can take: Flipping a neighboring spin or applying an update to the spin configuration.
	}
\end{figure}

The actions that the agent located at spin $\sigma$ can execute on the environment contains two types of operations:  Select and flip a neighboring spin  $\sigma_i, i \in \{0,..,5\}$,  $a_i$, and  propose an update, $a_{\text{update}}$ (Fig.~\ref{fig:local}).
At each step, the agent is located  at a spin and execute one action from the action space according to the policy $\pi_\theta(a|s)=[p(a_0), \ldots, p(a_5), p(a_{\textrm{update}})]$,  the probability distribution of taking an action $a$ for a given state $s$.
We note that all six neighbors are available in our RL framework, while in the conventional loop algorithm only two are allowed. 
For example, the policy for the loop  algorithm corresponding to the local environment shown in Fig.~\ref{fig:local} is 
$ \pi_{\text{loop}} = [0, 0, 0, 0, 1/2, 1/2, 0]$ 
since only the two outward pointing spins can be chosen with equal probability.
Therefore, we design the actions in a less restrictive way and allow the agent  to learn the effective trajectory by exploring the physical model.
In addition, the action $a_{\text{update}}$ provides  flexibility for the agent to terminate  at any step. 
As we will show in the following, after sufficient exploration of the physical environment, the agent tends to perform $a_{\text{update}}$ only when a loop is formed.

\begin{figure}
  \centering
  \includegraphics[width=\columnwidth]{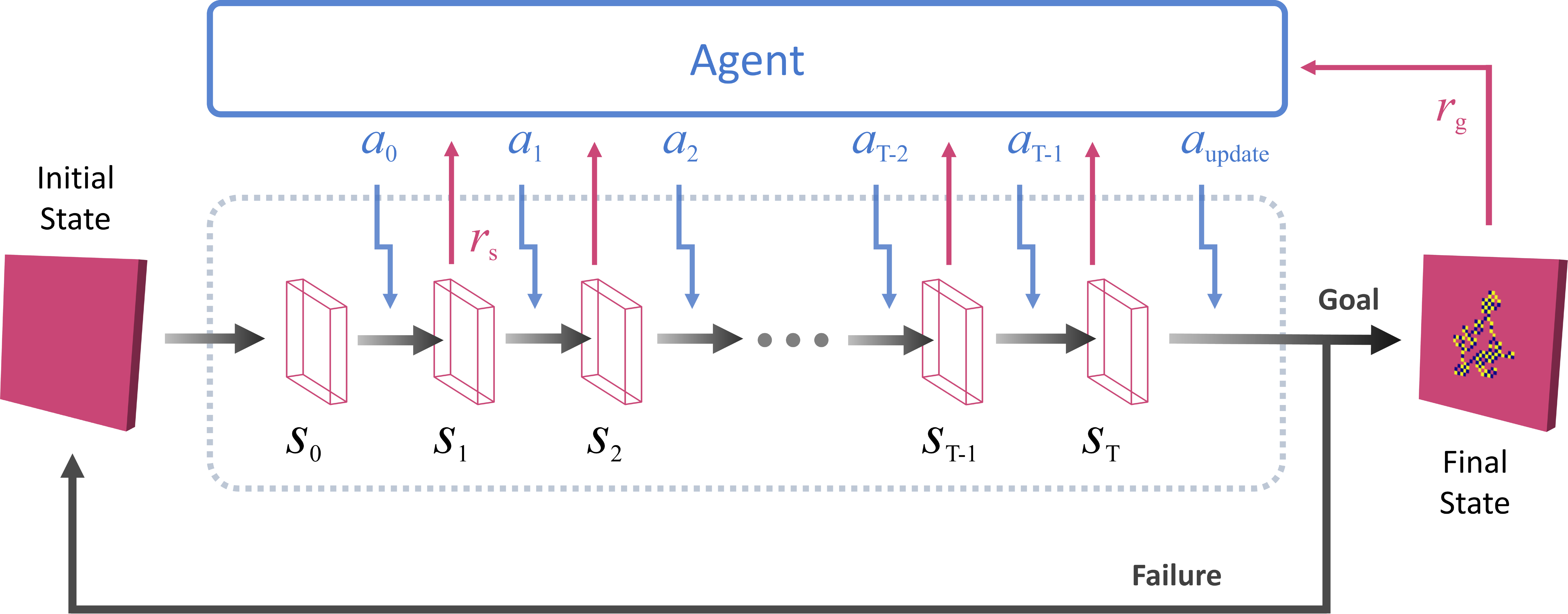}
  \caption{\label{fig:interface} (Color online)
    Architecture of agent-environment interface. 
    From an initial  ice state, the agent keeps executing actions on the environment and receiving corresponding rewards until the agent decides to make an update and finishes the episode. 
    In the final configuration, only the spins altered by the agent are depicted.
  }
\end{figure}

The objective of RL is maximizing the cumulative rewards. 
In playing Atari games, the game score is an obvious choice for reward. 
In our application, however, searching for update patterns hidden in the environment does not translate directly into an optimization problem. 
Therefore, the reward function  serves as the guiding principal to inform the agent whether a new ice state is generated.
The restricted environment itself would be the main reason that versatile policies emerge during the training process~\cite{heess2017emergence}.
We design the sparse reward function as a step function given at the end of each episode, 
\begin{align*}
  r_T(s,a) = \left\{
    \begin{array}{ll}
      r_g = 1 \text{, if the final state is an ice state} \\
      r_f = 0 \text{, otherwise} \\
    \end{array}
    \right.
\end{align*}
Once the action $a_{\text{update}}$ is executed, the episode will be terminated and the agent will receive the reward $r_T$. 
This encourages the agent to perform $a_{\text{update}}$ only when the environment is in a defect-vacuum state and thus generate a closed loop in each episode. 
Since the closed-loop condition is relatively rare, at most of the steps the agent receives no reward signals.
This kind of approximately silent environment usually causes inefficient sampling and vanishing gradient in the training process. 
To avoid this problem, we also define a stepwise reward $r_s(s,a)$  to encourage the agent to explore the environment and attempt to generate larger loops. 
The stepwise rewards are assigned at each step regardless of the action taken by the agent and chosen to  be relative small such that the agent recognizes $r_g$ as the ultimate goal~\cite{ng1999policy}. 
In practice, we keep the ratio of the stepwise reward to the goal reward as $r_{s}/r_{g}\sim \mathcal{O}(N^{-1})$, where $N$ is the number of sites of the system.

\begin{figure}[tb]
  \centering
  \includegraphics[width=\columnwidth]{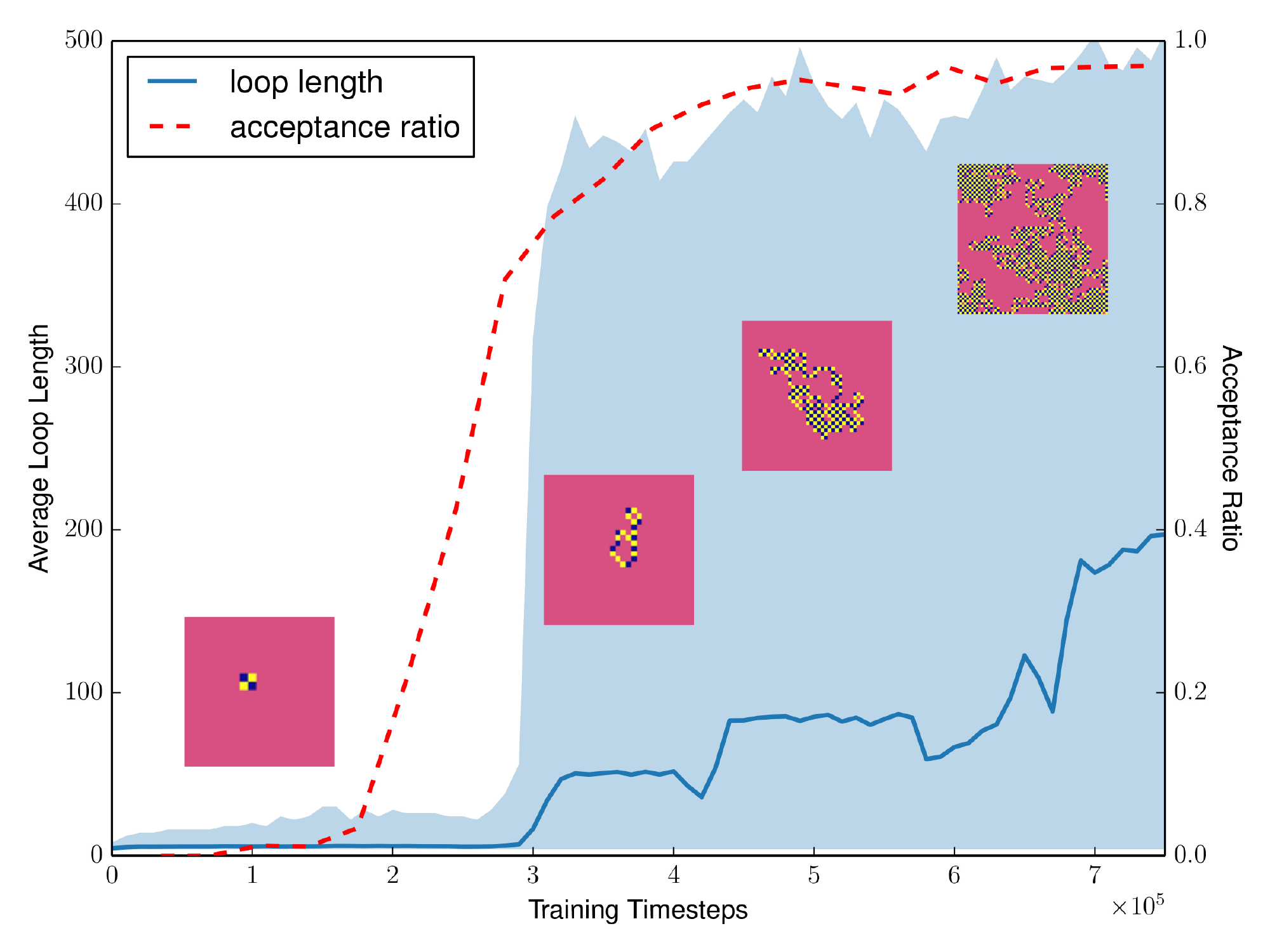}
  \caption{
    \label{fig:learning}
   (Color online) Agent training process for $L=16$.
   The shaded area presents the loop length distribution, bounded by maximum and minimum lengths.
   The dashed line represents the acceptance ratio of the agent's proposed move.
    At the early stage of learning, the agent only creates $2\times2$ smallest loops to obtain the goal rewards. 
    In this phase, agent learns appropriate timing to execute $a_{\text{update}}$.
    Based on the learned small-loop strategy, the agent realizes that ice rule allows for larger loops to occur if there are no more defects are generated in the process.
    After $7\times 10^{5}$ training steps, the acceptance ratio of the agent-proposed moves reaches close to 100\%.
  }
\end{figure}

\section{Training Process and Agent Behavior\label{training}}

At the beginning of the training process, the agent lacks  any prior knowledge about the physical environment and performs random moves to collect data.
After a few trials, the agent quickly gains the ability to generate a  $2\times 2$ square loop, the smallest closed loop providing a non-zero goal reward. 
In the absence of stepwise reward, the agent would exploit this policy by proposing square loops frequently. 
However, the stepwise reward encourages the agent to explore the possibilities beyond the square loop. 
The distribution of loop length as a function of training time is shown in \cref{fig:learning}. 
After about $3\times 10^{5}$ training episodes, the agent  becomes proficient at performing larger random loops and executes the update action correctly. 
While the maximum loop length shows a clear jump at this point, the average loop size grows steadily.
The acceptance ratio of the proposed moves also grows steadily during training, and approaches 100\% at later stage of the training.

\begin{figure}[tb]
  \centering
  \includegraphics[width=\columnwidth]{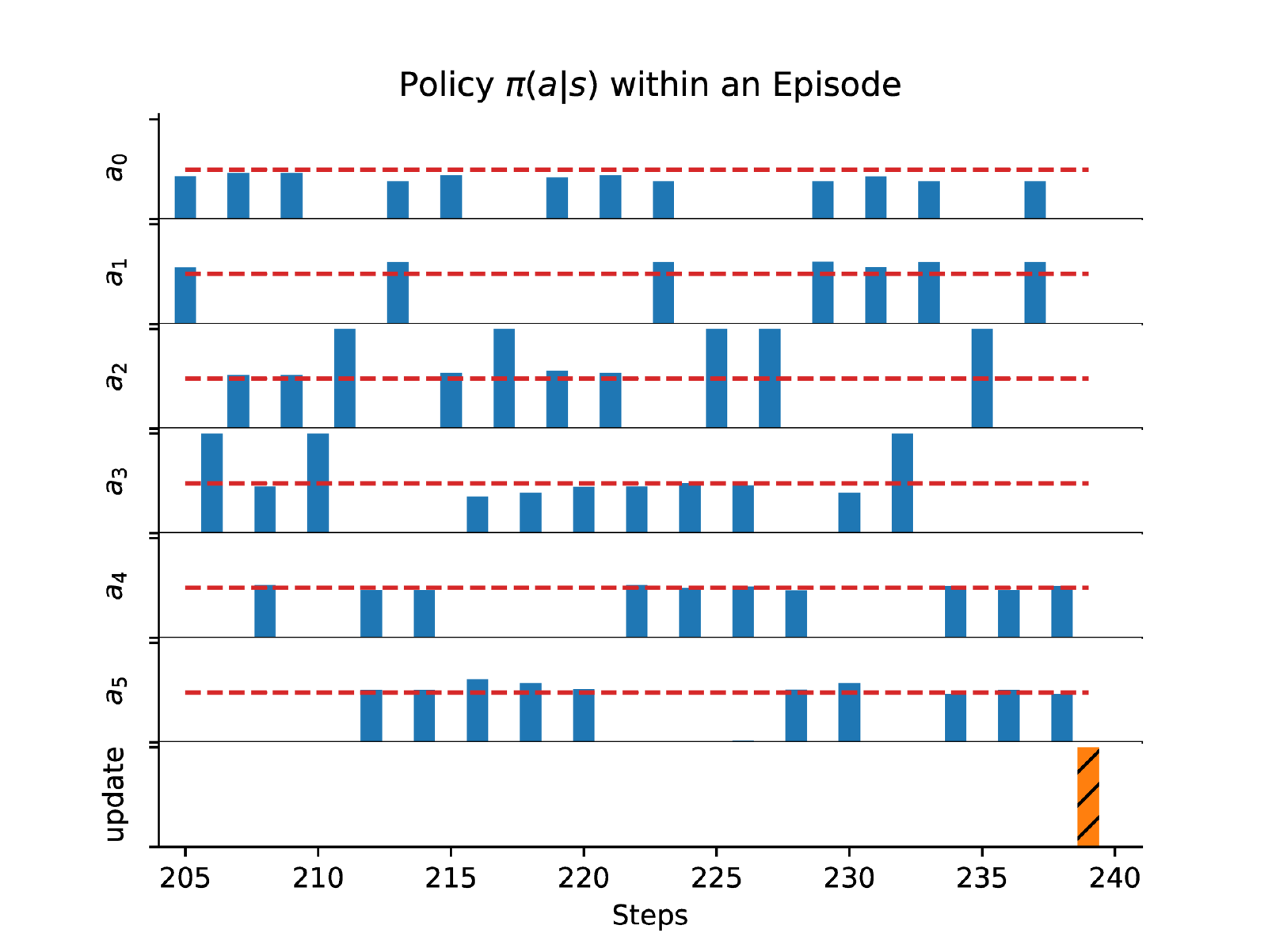}
  \caption{
    \label{fig:policy}
    (Color online) Decision-making process of the agent illustrated by the policy $\pi(a|s)$. 
    Most of the time, the  agent behaves in a similar way to conventional long-loop algorithm which makes stochastic selection from two outgoing spins (with probability $0.5$, denoted by red dashed lines). 
    When the trajectory forms a closed loop, the agent executes the update operation with full confidence (orange striped bar).
  }
\end{figure}

\begin{figure}
  \centering
  \includegraphics[width=\columnwidth]{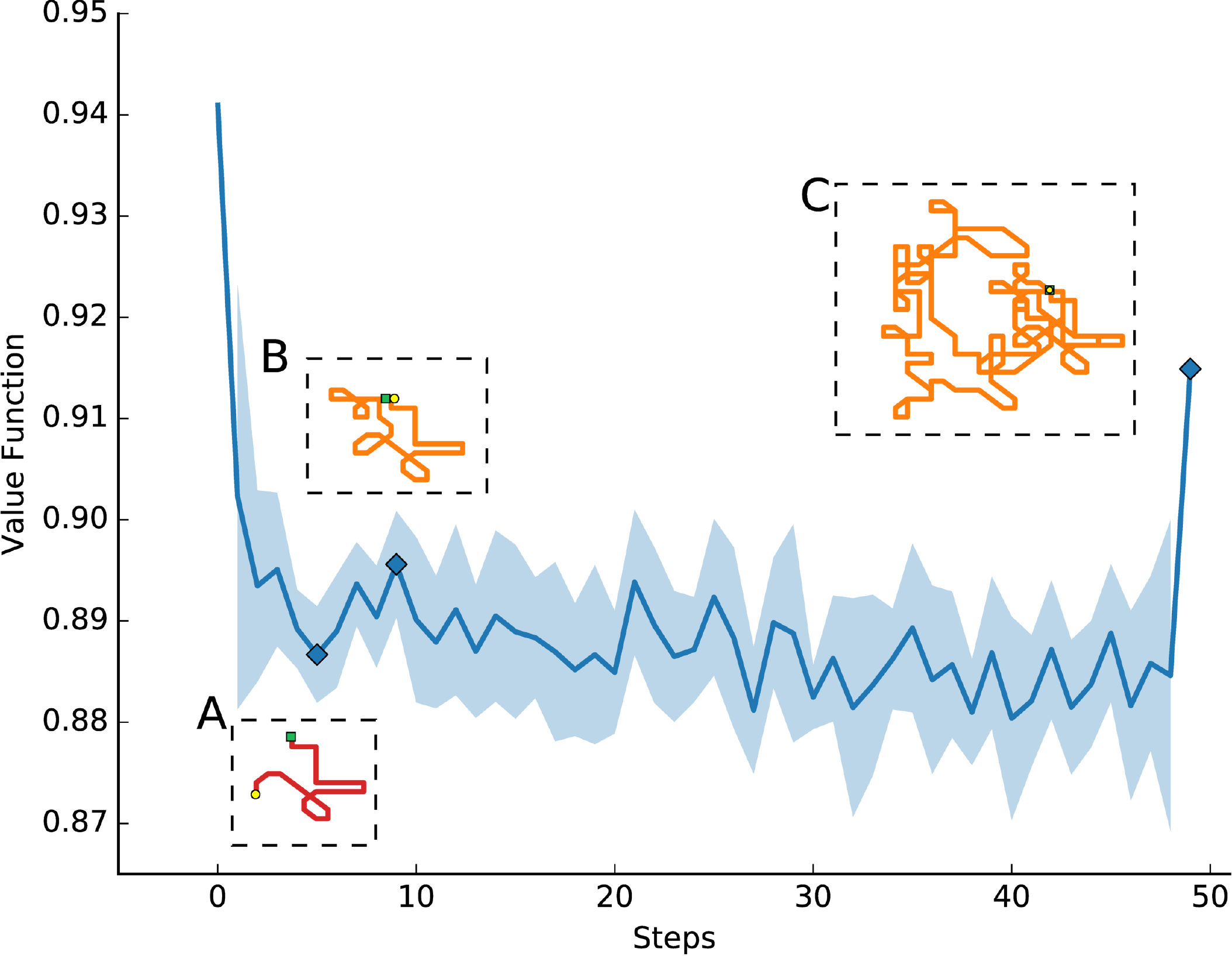}
  \caption{
    \label{fig:valfunc}  
    (Color online) State-value function $V^\pi(s)$ shows the expected future reward  in the process of a loop formation.
    Shaded area shows the distribution of the value, bounded by  the maximum and minimum value.     
      A local minimum occurs when endpoints (circle and square) are far apart (A), and  a local maximum occurs when  two endpoints move closer to each other (B). 
    When  a closed loop emerges, the value function shows a jump and large reward is expected (C).
  }
\end{figure}

In order to generate a new ice configuration, the agent needs to learn to annihilate the pair of defects by closing the loop. 
The well-trained agent can be regarded as a policy-guided machine in which the policy $\pi(a|s)$ does the decision-making of local spin flipping and the state-value function $V(s)$ serves as the  loop-closing detector.
In \cref{fig:policy}, we present the decision-making process by the agent near the end of one episode. 
In most of the steps, the policy distribution shows two possible candidate spins with almost equal probability. 
This behavior is similar to that of the conventional loop algorithm where one of the two out-going spins are selected with equal probabilities~\cite{Barkema:1998eo}. 
However, there exist time steps that the agent takes deterministic action to move in certain direction.
This indicates the agent is exploiting  knowledge acquired during the training.
Once the loop is closed, the agent decides to perform an update action ($a_{\text{update}}$) and terminate the episode with certainty.
This behavior is consistent with the loop closing condition in the loop algorithm.
We can also use the state-value function $V(s)$  as an indicator of the agent behavior since it implies the cumulative future reward.
In \cref{fig:valfunc} we show the evolution of $V(s)$ in the process  of a single loop formation. 
When the episode starts, the system is close to an ice state and the value function gives high expectation value. 
After a few steps of executing the local policy, the value function drops to  a local minimum as ends of the segment move  apart and going toward the opposite directions (inset A in Fig.~\ref{fig:valfunc}). 
During the exploration stage, it is possible that  two ends move closer, and the value function reaches a local maximum (inset B in Fig.~\ref{fig:valfunc}). 
At the final step, the value function jumps suddenly, anticipating the goal reward to be obtained by performing the update action (inset C in Fig.~\ref{fig:valfunc}).
This indicates that the state-value function plays the role of loop-closing detector and recognizes the global pattern of the environment.


\section{Sampling using trained policies\label{sampler}}

\begin{figure}[t]
\centering
\includegraphics[width=0.8\columnwidth]{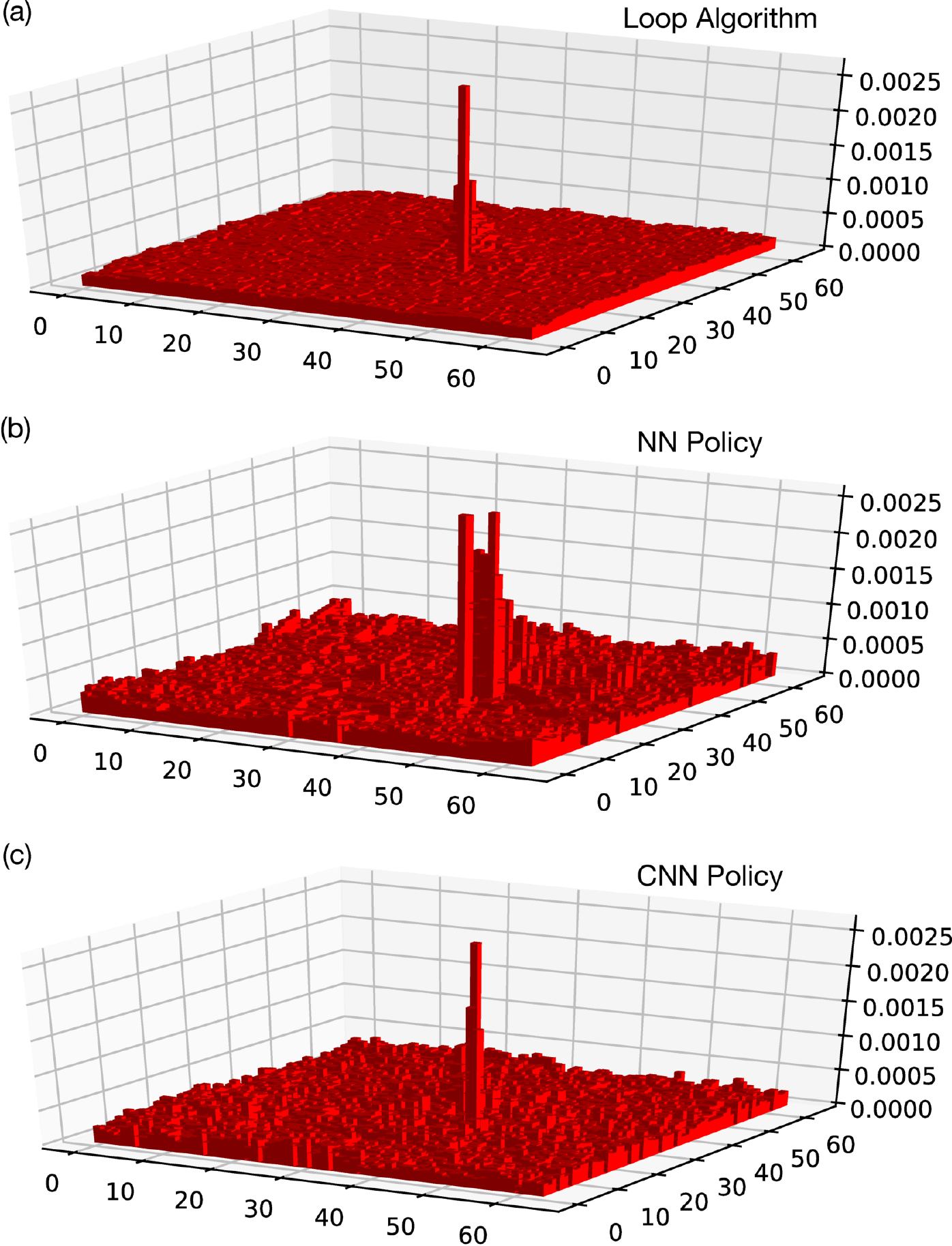} 
\caption{\label{fig:memory_samesite} (Color online) Probability that a site is visited starting from the same initial site and  ice configuration for (a) loop algorithm, (b) NN policy, and (c) CNN policy. 
The NN policy  shows significant memory effects while the CNN policy shows similar behavior as the loop algorithm. 
}
\end{figure}

As mention in Sec.~\ref{sec:agent}, two types of  policies are trained by supplying different sets of observations to the agent.
The NN policy is trained with only the local observation $O_\textrm{local}$, and the CNN policy is trained with both the local and global observations $O_\textrm{local}$ and $O_\textrm{global}$.
We will use these trained policies as samplers in the MCMC for the square ice model. 
With the capacity of  deep learning models, it is possible the network simply memorizes the configurations it generated in the learning process.
In order to check this, we start the trained agents from the same ice configuration at the same initial position to perform one update.
We expect that the agent's behavior should become independent of the initial starting position and explore the space without special preferences.
The baseline loop algorithm shows the probability of each site being visited  rapidly smears out (\cref{fig:memory_samesite}(a)) with the number of steps.
We observe, on the other hand,  strong memory effects in the NN policy (\cref{fig:memory_samesite}(b)) such that the agent tends to linger around the initial site; while the CNN policy behaves similar to the baseline loop algorithm (\cref{fig:memory_samesite}(c)). 

\begin{figure}[t]
  \centering
  \includegraphics[width=\columnwidth]{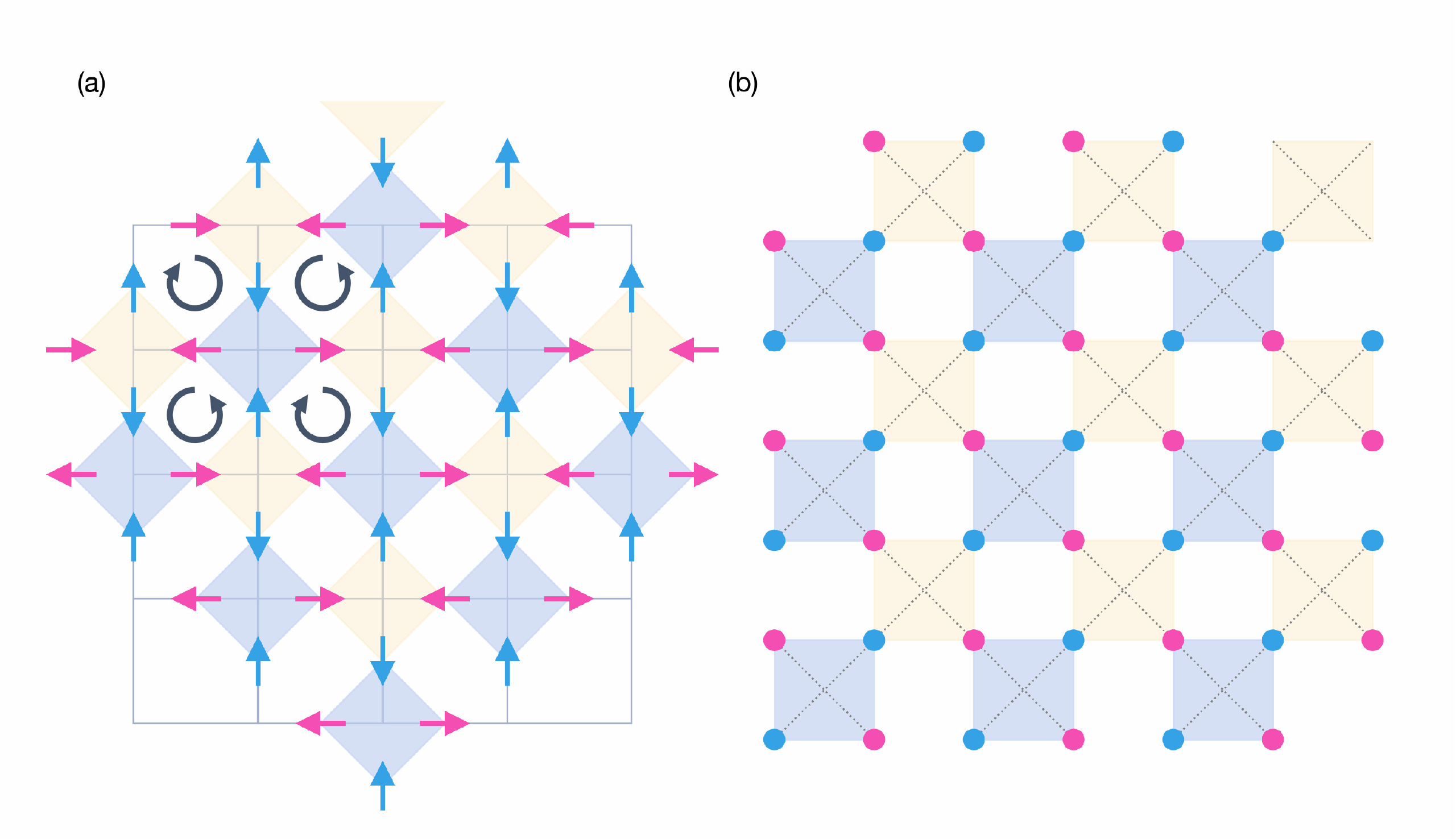}
  \caption{
    \label{fig:mapping} (Color online) Equivalence mapping between the (a) vertex spins on a square lattice and (b) Ising variables on a checkerboard lattice by a 45$^\circ$ rotation.
   This specific configuration corresponds to the ice state with a N\'{e}el order, where a regular vortex pattern (black circular arrows) is shown. 
  }
\end{figure}

\begin{figure*}[t]
\centering
	\includegraphics[width=1.0\linewidth]{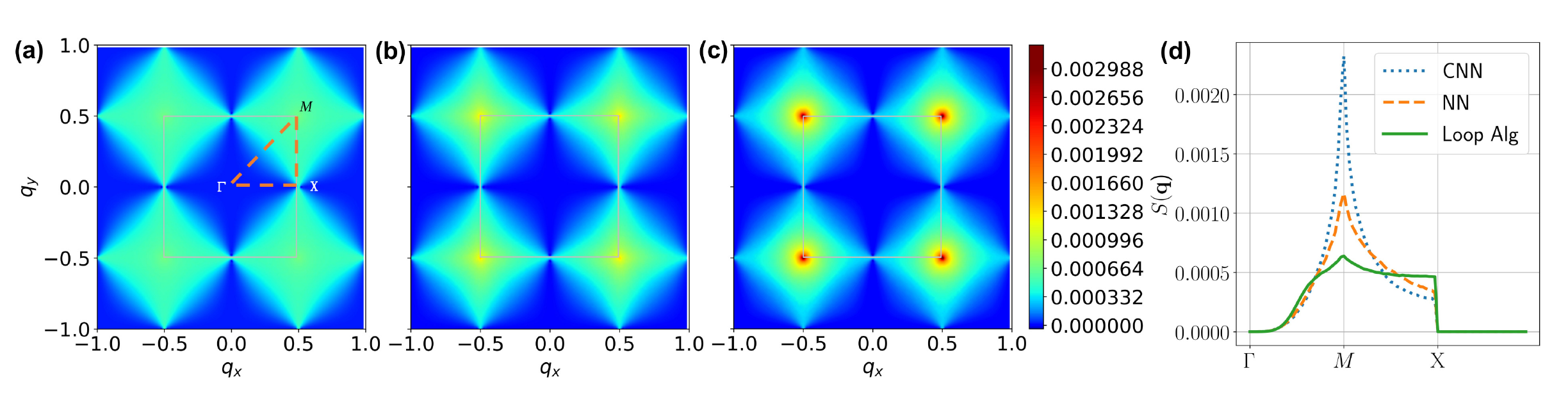} 
	\caption{	\label{fig:Sq} (Color online) The structure factor for $L=64$ computed using the configurations generated from (a) the conventional loop algorithm, (b)  the  NN policy, and (c) the CNN policy. 
	The gray square indicates the magnetic first Brillouin zone. 
	The NN and CNN policies are trained in a $L=32$ setting and a moving horizon embedding described in the Appendix A is used to generate states for $L=64$.
	(d) Line cuts of $S(\mathbf{q})$ along high symmetry directions in the first Brillouin zone as indicated in the orange (dashed) lines in (a).
 Both the trained policies show bias toward the N\'{e}el order.}
\end{figure*}

We now use the trained policies to propose samples in the Markov chain Monte Carlo to compute physical observables. 
In particular, we are interested in  correlation functions such as the magnetic structure factor defined as  
\begin{align}
S(\mathbf{q}) &=\frac{1}{L^2} \left\langle s_\mathbf{q} s_{-\mathbf{q}} \right\rangle, \mbox{where } s_\mathbf{q} = \sum_r s_{\mathbf{r}} e^{-i \mathbf{q} \cdot \mathbf{r}},
\end{align}
where the  Ising variables $s_{\mathbf{r}}$ on a checkerboard lattice are obtained by an equivalence mapping from the vertex spins on a square lattice~(Fig.~\ref{fig:mapping}), and $\mathbf{q}$ is the momentum.
In order to reproduce the correct correlation, the configurations should be sampled according to the underlying distribution.
All  ice configurations  in the square ice model are equally probable; that is, all ice configurations should be sampled equally for the sampling to be ergodic. 
Again, we use the structure factor generated by the loop algorithm as our baseline. 
Fig.~\ref{fig:Sq} shows the spin structure factors using samples generated by different algorithms.
Both the NN (Fig.~\ref{fig:Sq}(b)) and CNN policies (Fig.~\ref{fig:Sq}(c)) can capture the diffuse scattering present in the model; however, it is clear that correlation is biased for the trained policies.
We plot  in Fig.~\ref{fig:Sq}(d) the structure factor  $S(\mathbf{q})$ along the high symmetry direction in the first Brillouin zone. 
We find that some weight of the diffusing scattering is missing in the trained policies,  and spectral weight is shifted  toward the $\mathbf{Q}=(\pi/2,\pi/2)$ point, indicating bias toward the N\'{e}el ordered states for the trained policies.       
Surprisingly the CNN policy is biased more toward the N\'{e}el states than the NN policy, although the former has less memory effects due to the global view. 
In the N\'{e}el order, spins  arrange in  a regular vortex pattern (Fig.~\ref{fig:mapping}). 
This pattern requires a global view to establish; thus, the CNN policy tends to bias more toward the  N\'{e}el states.

Clearly, the trained policies are not sampling all the ice states equally, and the ergodicity is broken. 
This is related to the fact that the machine policy contains actions that makes deterministic moves. 
These actions indicate that agent somehow memorizes special spin configurations during the training process. 
Therefore, directly using the policies as samplers introduces bias in the sampling.
This is the consequence of not enforcing the detailed balance condition during the training, but may be corrected if one obtains an estimate of the density of the bias in the transition probability.
We note this should not be considered a failure of the current framework since the probability distribution of ice states is  multimodal with equal weights, which posts a significant challenge to properly sample. 
On the other hand, by adjusting the stepwise reward, it is possible to bias the agent to create larger loops. 
Therefore, this can be used to  as an overrelaxation step to decorrelate  samples in the Markov chain, when combined with the  conventional loop algorithm.

\section{Conclusion\label{conclusion}}

In this work, we develop a framework using deep reinforcement learning  to generate topologically constrained ice states. 
We successfully train an agent to propose stepwise actions that when combined together can transit from one ice state to another. 
By rewarding the agent when a correct ice state is generated, several physical insights emerge out of the training.
Although the  ice rule  is not given explicitly, the agent learns to move in a path without creating additional defects. 
The agent also learns  to distinguish open and closed loops, and perform the update action when the loop closes. 
The physical constraints are built  into the reward function, not the policy itself.
By using deep reinforcement learning, we show that the machine can actually learn the global loop pattern and propose updates without prior knowledge of the ice rule.
Therefore, it is possible to extend this framework to other physical models. 
For example, quantum Monte Carlo simulations of  quantum spin ice models~\cite{Henry:2014vh,Wu:2019it,Wang:2018tt}, toric code and related gauge models~\cite{Trebst:2007uh,Tupitsyn:2010lh,Kamiya:2015ls} have been  difficult due to the physical constraints present in the model.
The generality of the RL framework presented here can explore an enlarged state space and potentially discover new  sampling schemes through the automatic exploration of the machine agent on a constrained model.
These update proposals can be used as candidate policies in the PGMC~\cite{pgmc2018}  where  detailed balance  can be satisfied. 
It would also be interesting to model the policy using neural networks with normalizing flows~\cite{Dinh:2014rz,Jimenez-Rezende:2015kz,Dinh:2016ng,Papamakarios:2017zx,Li:2018er}.
Combining the exploration capability of RL with the  unbiased control of the training and inference of the reversible policy model can potentially lead to the discovery of efficient MCMC algorithms without detailed balance~\cite{Suwa:2010qv} for  constrained models.  
Finally, we note that the training of the model takes roughly three days on a Xeon workstation, and each inference step of the CNN policy takes about $0.75$~ms. 
Both the conventional loop algorithm and the CNN policy  take about five minutes to generate 1000 loops on a $L=32$ system. 

\section{Acknowledgments}
Y.-J.K. thanks T. Bojesen for useful discussion and critical reading of the manuscript. 
This work was supported  in part by the Ministry of Science and Technology (MOST) of Taiwan under Grants No. 105-2112-M-002-023-MY3, and 107-2112-M-002 -016 -MY3 , 108-2918-I-002 -032, and  by the National Science Foundation under Grant No. NSF PHY-1748958. 
We are also grateful to the National Center for High-performance Computing
for computer time and facilities. 

\appendix

\section{Sliding horizon scheme}
We train the agent policy in a $L_t\times L_t$ square lattice with the periodic boundary condition. 
In general, the trained network can only process fixed input dimension  assigned before training.
In order to use the agent in a larger system, we exploit the translational invariance of the policy and use the sliding horizon embedding that dynamically crops the input image and moves the horizon together with the agent. 
The observation information within the $L_t\times L_t$  window is provided to the agent. 
When the agent reaches the boundary of the  window, the window is recentered around the agent as the agent traversing the $L\times L$ lattice. 
This scheme provides the agent  the view similar to the training stage.
It also allows  us to access  system sizes larger than the training size  as presented in the main text.

\begin{figure}[t] 
  \centering
  \includegraphics[width=0.6\columnwidth]{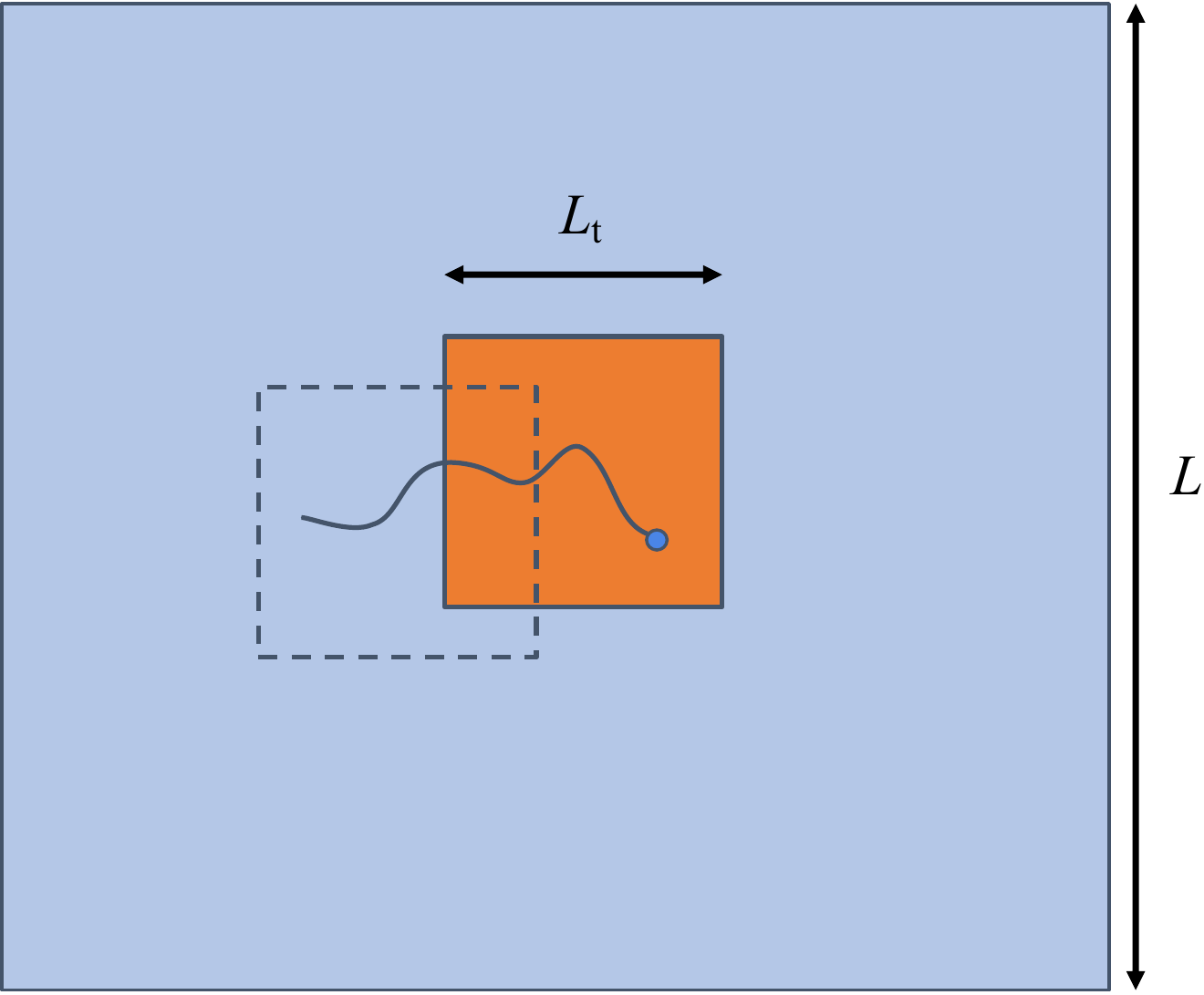}
  \caption{
    \label{fig:moving}  
	(Color online) The sliding horizon embedding scheme. 
	 The trained agent (blue circle) takes input from an $L_t\times L_t$ window (dark inner square) embedded in a $L\times L$ lattice. 
	 Dashed square corresponds to the initial position of the observation window. 
	 }
\end{figure}

\section{Network architecture and hyperparameters\label{app:network}}

Here we present  details of the specifications  of the neural network architecture  and the hyperparameters used in this paper.
Table~\ref{tb:network} lists the specification of the local and multi-channel networks for the agent. 
Table~\ref{tb:hparam} lists all of hyperparameters, system settings and environment configurations for future experiment reproductions.
\begin{table}[t]
  \caption{\label{tb:network} Neural network architectures for the agent.} 
  \centering
  \begin{tabular}{ |l|l|l| }
    \hline
    Name & Layer & Size \\ 
    \hline
    \multirow{4}{*}{Local Network, NN Policy} 
      & linear (relu) & $32$ \\
      & linear (relu) & $64$ \\
      & linear (relu) & $128$ \\
      & parallel (relu) & $7$, $1$ \\
      \hline

    \multirow{7}{*}{Multi-channel Network, CNN Policy}
      & linear (relu) & $32$ \\
      & linear (relu) & $64$ \\
      & conv   (relu) & $32, 3\times 3$, stride $2$ \\
      & conv   (relu) & $16, 3\times 3$, stride $2$ \\
      & concat & $1088$ \\
      & linear (relu) & $128$ \\
      & parallel (relu) & $7$, $1$ \\
      \hline
  \end{tabular} 
\end{table}

\begin{table}[t]
\centering
\caption{\label{tb:hparam} Hyperparameters used in  training. } 
\begin{tabular}{ |l|l|l| }
  \hline
  Type & Hyperparameter & Value \\ \hline
  \multirow{9}{*}{Training} 
    & Solver type & ADAM \\
    & Learning rate & $10^{-4}$ \\
    & $\beta_1, \beta_2$ (hparam for ADAM) & $0.9, 0.999 $ (default) \\
    & Total steps & $10^{9}$ \\
    & Update steps & $120$ \\
    & Entropy regularization, $\lambda$ & $0.0001$ \\
    & Exploration & $\epsilon-$greedy, $1\%$ \\
    & Gradient clip & $ 40.0 $ \\
    & Discount factor $\gamma$ & $ 0.99 $ \\
    \hline

  \multirow{5}{*}{System}
    & Number of parameter server & 1 (host)\\
    & Number of workers & 8 \\
    & Number of policy monitor & 1 \\
    & TensorBoard process & 1 \\
    & Evaluation duration & 60 seconds \\
    \hline

  \multirow{9}{*}{Environment} 
    & Lattice size & $L=16$ \\
    & Observation shape & $O_g : 32\times 32\times 1$, $O_l: 10$ \\
    & Action shape & $7$ \\
    & Stepwise reward & $+0.001$ \\
    & Target reward & $+1.0$ \\
    & Timeout steps & $ 1024$ \\
  \hline
\end{tabular}
\end{table}

\bibliography{refs.bib}
\end{document}